\documentclass{ws-procs9x6}

\begin{document}
\title{Calibration Studies and the Investigation of Track Segments within
Showers with an Imaging Hadronic Calorimeter}
\author{Shaojun Lu, on behalf of the CALICE Collaboration}
\address{Max-Planck-Institute for Physics, and Excellence Cluster Universe,
Munich, Germany}
\begin{abstract}
Abstract-The CALICE collaboration has constructed a highly granular hadronic
sampling calorimeter prototype with small scintillator tiles individually read
out by silicon photomultipliers (SiPM) to evaluate technologies for the ILC.
The imaging capability of the detector allows detailed studies of the
substructure of hadronic events, such as the identification of minimum
ionizing track segments within the hadronic shower. These track segments are
of high quality, so that they can be used for calibration, as an additional
tool to Muons and to the built-in LED system used to monitor the SiPMs. 
These
track segments also help to constrain hadronic shower models used in Geant4.
Detailed MC studies with a realistic model of an ILC detector were performed
to study the calibration requirements of a complete calorimeter system. The
calibration strategy was tested on real data by transporting calibration
constants from a Fermilab beam test to data recorded at CERN under different
conditions.
\end{abstract}
\bodymatter
\section{The CALICE Detectors}

The goal of the CALICE experimental program is to establish novel technologies
for calorimetry in future collider experiments and to record electromagnetic
and hadronic shower data with unprecedented three dimensional spatial
resolution for the validation of simulation codes and for the test and
development of reconstruction algorithms. Such highly granular calorimeters
are necessary to achieve an unprecedented jet energy resolution at the
International Linear Collider \cite{JBra} using particle flow algorithms
\cite{MTho}. The CALICE analog hadron calorimeter (AHCAL) prototype is a 38
layers sampling calorimeter, which is built out of scintillator tiles with
sizes ranging from $30 \times 30$mm$^2$ in the core of the detector to $120
\times 120$mm$^2$ . The light in each scintillator cell is collected by a
wavelength shifting fiber, which is coupled to a silicon photomultiplier
(SiPM) SiPM. The SiPMs are produced by the MEPhI/PULSAR group \cite{GBon}.
They have a photo-sensitive area of $1.1 \times 1.1$mm$^2$ containing 1156
pixels with a size of $32 \times 32 {\mu}$m$^2$. In total, the calorimeter has
7608 channels. A built-in LED calibration system with UV LEDs, is coupled to
each cell by clear fibers, and equipped with PIN diodes to monitor the LED
light intensity. The performance of the AHCAL was validated with positrons at
various energies. Good linearity and satisfactory agreement with simulation
up to an energy of 50 GeV has been observed. \cite{CAN014}

\section{Track segments in hadronic showers}

The high granularity of the active layers in the hadronic calorimeter and the
cell-by-cell readout gives the CALICE detectors unprecedented imaging
capabilities. This is exploited to study the topology of hadronic events in
detail. Track segments created by charged particles produced within the
hadronic  shower can be identified, provided the particles travel an
appreciable distance before interacting again and are separated from other
activity in the detector. 

\begin{figure}[!htb]
\begin{center}
       \psfig{file=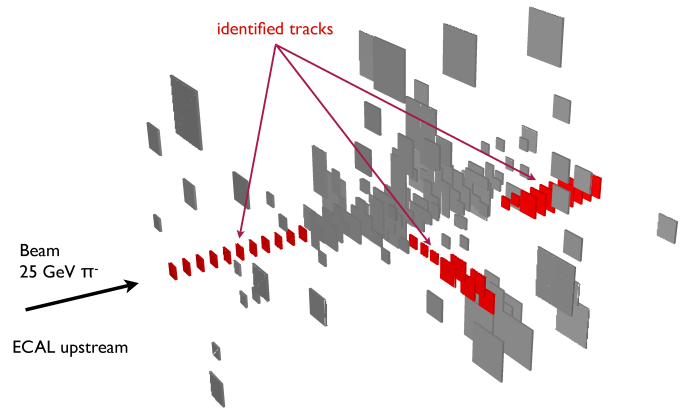,width=0.8\textwidth}
\end{center}
   \caption {Identification of minimum-ionizing track segments within
   hadronic showers.}
   \label{fig:HCALTrackSegments}
\end{figure}

Figure \ref{fig:HCALTrackSegments} demonstrated that three
tracks were found within one 25 GeV $\pi^-$ event. The properties of these
tracks are sensitive to the substructure of the hadronic shower, and can thus
serve as a powerful probe for hadronic shower models.  The track segments
identified in hadronic showers have a high quality, and are suitable for
detector calibrations via the extraction of the most probable value of the
energy loss in each cell along the track. In the CALICE detector prototype,
this technique was used to study the temperature dependence of the detector
response \cite{FSi1}.

\section{In-situ calibration strategy and requirements for an ILC calorimeter}

Due to the underground location, the orientation of the detector layers, the
power pulsing of the front end electronics, and due to the high granularity,
calibrating a hadronic calorimeter with approximately 8 million channels at a
future ILC detector is a significant challenge. Cosmic rays might not be
sufficient for monitoring the energy scale in-situ. The track segments can be
used for calibration purposes. A simulation study performed with a model of a
complete ILC detector showed that a layer by layer calibration to a precision
of better than 5\% can be achieved with an integrated luminosity of 10
pb$^{-1}$ at the Z$^0$ resonance.

\begin{figure}[!htb]
\begin{center}
       \psfig{file=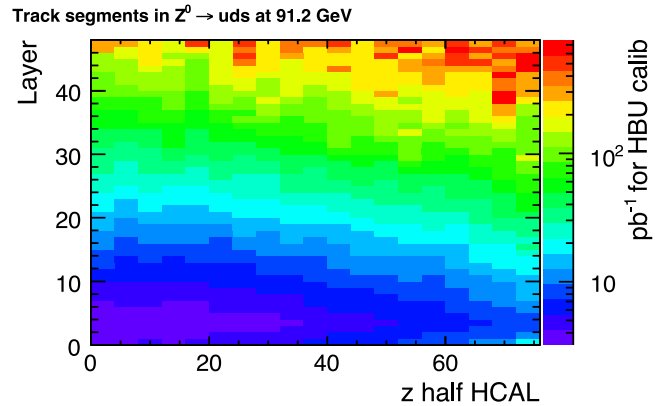,width=0.8\textwidth}
\end{center}
   \caption{Required luminosity for 1000 identified tracks in
   hadronic events per electronic unit HBU at $Z^0$ resonance running in the
   barrel AHCAL. One half of the AHCAL is shown, with the z coordinate of the
   interaction point at the lower left corner, z is measured here in units of
   tiles (3 cm).}
   \label{fig:Calibration}
\end{figure}

Figure \ref{fig:Calibration} shows the
integrated luminosity required for the calibration of individual electronic
subunits (144 cells), and illustrates the distribution of the number of found
tracks within the calorimeter volume. At 500 GeV, this can be achieved with two
weeks worth of data, for a calibration of the first 20 layers, one day is
sufficienct.  

The single photon resolution of the SiPMs, together with an
LED light system, can be used to perform a gain calibration of the photon
sensor, and to monitor changes due to short term variations of environmental
parameters. For a channel-by-channel intercalibration of the
complete cell response, particles are necessary. Before installation, such a
calibration can be performed in a beam, analogous to the muon calibration
currently used for the CALICE detector system.\cite{FSi2} Monitoring of long
term variations, such as changes in the light yield of the scintillator have
then to be performed in-situ. This can be achieved with track segments which
are identified within the hadronic showers. To determine the requirement of
calibration accuracy for the calorimeter of an ILC detector, a study has been
preformed for the scintillation-tile hadron calorimeter, based on full 
detector simulations. These studies demonstrate that spreads up to 10\% in
the layer to layer and cell by cell intercalibrations do not lead to a
noticeable degradation of the overall performance in terms of the diijet energy
resolution with particle flow algorithms.

\section{Test of calibration strategy on real data}

The calibration strategy was tested on real data by transporting calibration
constants from a Fermilab beam test to data recorded at CERN. Two ways of
transporting have been used, one based on the known temperature and voltage
dependence of the SiPM response, and one based on measured changes in gain.

\begin{figure}[!htb]
\begin{center}
       \psfig{file=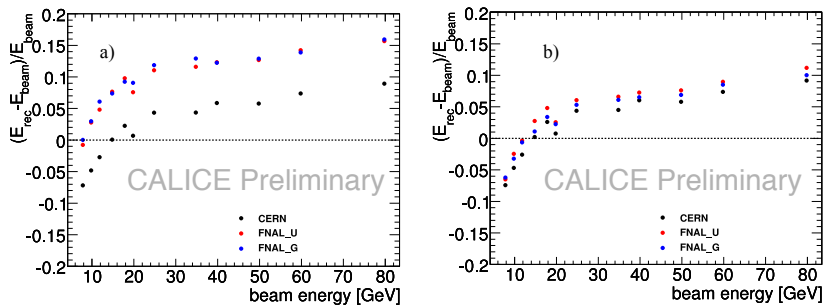,width=1.0\textwidth}
\end{center}
   \caption{a) Residual from linearity of reconstructed hadron showers fully
   contained in the AHCAL, in the range 8 to 80 GeV.
b) Residual from linearity of
   reconstructed hadron showers fully contained in the AHCAL, in the range 8 to
   80 GeV. adding the in-situ MIP stub correction layer-by-layer to each of the
   sample.} 
   \label{fig:combin2}
\end{figure}

Figure \ref{fig:combin2} a shows a set of pion runs taken at CERN in the
energy range 8 to 80 GeV were analyzed. The residual from linearity for the
reference CERN calibration and for the two transported calibration sets were calculated.
The hadronic energies calibrated with FNAL transported coefficients are
approximately 5\% higher than the reference CERN calibration sample. The two
transport methods are in agreement with each other. The nonlinearity of the
detector response is due to the requirement of full containment. After
applying to all energy points an unique layer-by-layer correction derived from
the identified track segments, a clear improvement in the agreement between
different calibration methods is observed in figure \ref{fig:combin2} b. This
demonstrates that an intercalibration of the detector modules in a beam before
installation, and a layer-wise calibration with track segments after
installation using regular physics data is sufficient to guarantee stable
performance of the calorimeter at a future linear collider.

\end{document}